\documentclass[aps,prl,twocolumn,amsmath,superscriptaddress]{revtex4-1}  
\usepackage{bbm}
\usepackage{mathrsfs}
\usepackage{amsmath}
\usepackage{amsfonts}
\usepackage[colorlinks=true,citecolor=blue,anchorcolor=blue]{hyperref}
\usepackage{graphicx,epstopdf}
\usepackage{float}
\usepackage{subfigure}
\usepackage{epsfig}
\usepackage{dcolumn}
\usepackage{bm}
\usepackage{color}
\usepackage{natbib}
\usepackage{amssymb}
\usepackage{xcolor}
\usepackage{braket}
\begin{document}
\title{Large band-splitting in $g$-wave type altermagnet CrSb}

\author{Jianyang Ding}
\thanks{Equal contributions}
\affiliation{National Synchrotron Radiation Laboratory, University of Science and Technology of China, Hefei, 230026, China}
\affiliation{Shanghai Synchrotron Radiation Facility, Shanghai Advanced Research Institute,
Chinese Academy of Sciences, Shanghai 201210, China}
\affiliation{National Key Laboratory of Materials for Integrated Circuits, Shanghai Institute of Microsystem and Information Technology, Chinese Academy of Sciences, Shanghai 200050, China}

\author{Zhicheng Jiang}
\thanks{Equal contributions}
\affiliation{National Synchrotron Radiation Laboratory and School of Nuclear Science and Technology, University of Science and Technology of China, Hefei, 230026, China}

\author{Xiuhua Chen}
\thanks{Equal contributions}
\affiliation{School of Emerging Technology, University of Science and Technology of China, Hefei 230026, China}

\author{Zicheng Tao}
\thanks{Equal contributions}
\affiliation{ShanghaiTech Laboratory for Topological Physics $\&$ School of Physical Science and Technology, ShanghaiTech University, Shanghai 201210, China}

\author{Zhengtai Liu}
\email{liuzt@sari.ac.cn}
\affiliation{Shanghai Synchrotron Radiation Facility, Shanghai Advanced Research Institute,
Chinese Academy of Sciences, Shanghai 201210, China}
\affiliation{National Key Laboratory of Materials for Integrated Circuits, Shanghai Institute of Microsystem and Information Technology, Chinese Academy of Sciences, Shanghai 200050, China}

\author{Tongrui Li}
\affiliation{National Synchrotron Radiation Laboratory and School of Nuclear Science and Technology, University of Science and Technology of China, Hefei, 230026, China}

\author{Jishan Liu}
\affiliation{Shanghai Synchrotron Radiation Facility, Shanghai Advanced Research Institute,
Chinese Academy of Sciences, Shanghai 201210, China}
\affiliation{National Key Laboratory of Materials for Integrated Circuits, Shanghai Institute of Microsystem and Information Technology, Chinese Academy of Sciences, Shanghai 200050, China}

\author{Jianping Sun}
\affiliation{Beijing National Laboratory for Condensed Matter Physics and Institute of Physics, Chinese Academy of Sciences, Beijing 100190, China}
\affiliation{School of Physical Sciences, University of Chinese Academy of Sciences, Beijing 100190, China}

\author{Jinguang Cheng}
\affiliation{Beijing National Laboratory for Condensed Matter Physics and Institute of Physics, Chinese Academy of Sciences, Beijing 100190, China}
\affiliation{School of Physical Sciences, University of Chinese Academy of Sciences, Beijing 100190, China}

\author{Jiayu Liu}
\affiliation{National Key Laboratory of Materials for Integrated Circuits, Shanghai Institute of Mic rosystem and Information Technology, Chinese Academy of Sciences, Shanghai 200050, China}

\author{Yichen Yang}
\affiliation{National Key Laboratory of Materials for Integrated Circuits, Shanghai Institute of Microsystem and Information Technology, Chinese Academy of Sciences, Shanghai 200050, China}

\author{Runfeng Zhang}
\affiliation{National Synchrotron Radiation Laboratory and School of Nuclear Science and Technology, University of Science and Technology of China, Hefei, 230026, China}

\author{Liwei Deng}
\affiliation{National Key Laboratory of Materials for Integrated Circuits, Shanghai Institute of Microsystem and Information Technology, Chinese Academy of Sciences, Shanghai 200050, China}

\author{Wenchuan Jing}
\affiliation{National Key Laboratory of Materials for Integrated Circuits, Shanghai Institute of Microsystem and Information Technology, Chinese Academy of Sciences, Shanghai 200050, China}

\author{Yu Huang}
\affiliation{National Key Laboratory of Materials for Integrated Circuits, Shanghai Institute of Microsystem and Information Technology, Chinese Academy of Sciences, Shanghai 200050, China}

\author{Yuming Shi}
\affiliation{National Key Laboratory of Materials for Integrated Circuits, Shanghai Institute of Microsystem and Information Technology, Chinese Academy of Sciences, Shanghai 200050, China}

\author{Mao Ye}
\affiliation{Shanghai Synchrotron Radiation Facility, Shanghai Advanced Research Institute,
Chinese Academy of Sciences, Shanghai 201210, China}
\affiliation{National Key Laboratory of Materials for Integrated Circuits, Shanghai Institute of Microsystem and Information Technology, Chinese Academy of Sciences, Shanghai 200050, China}

\author{Shan Qiao}
\affiliation{National Key Laboratory of Materials for Integrated Circuits, Shanghai Institute of Microsystem and Information Technology, Chinese Academy of Sciences, Shanghai 200050, China}

\author{Yilin Wang}
\email{yilinwang@ustc.edu.cn}
\affiliation{School of Emerging Technology, University of Science and Technology of China, Hefei 230026, China}
\affiliation{Hefei National Laboratory, University of Science and Technology of China, Hefei 230088, China}

\author{Yanfeng Guo}
\email{guoyf@shanghaitech.edu.cn}
\affiliation{ShanghaiTech Laboratory for Topological Physics $\&$ School of Physical Science and Technology, ShanghaiTech University, Shanghai 201210, China}

\author{Donglai Feng}
\affiliation{National Synchrotron Radiation Laboratory and School of Nuclear Science and Technology, University of Science and Technology of China, Hefei, 230026, China}
\affiliation{New Cornerstone Science Laboratory, University of Science and Technology of China, Hefei, 230026, China}
\affiliation{Collaborative Innovation Center of Advanced Microstructures, Nanjing, 210093, China}

\author{Dawei Shen}
\email{dwshen@ustc.edu.cn}
\affiliation{National Synchrotron Radiation Laboratory and School of Nuclear Science and Technology, University of Science and Technology of China, Hefei, 230026, China}

\begin{abstract}
Altermagnetism (AM), a newly discovered magnetic state, ingeniously integrates the properties of ferromagnetism and antiferromagnetism, representing a significant breakthrough in the field of magnetic materials. Despite experimental verification of some typical AM materials, such as MnTe and MnTe$_2$, the pursuit of AM materials that feature larger spin splitting and higher transition temperature is still essential. Here, our research focuses on CrSb, which possesses N{\'e}el temperature of up to 700K and giant spin splitting near the Fermi level ($E_F$). Utilizing high-resolution angle-resolved photoemission spectroscopy and density functional theory calculations, we meticulously map the three-dimensional electronic structure of CrSb. Our photoemission spectroscopic results on both (0001) and (10$\overline{1}$0) cleavages of CrSb collaboratively reveal unprecedented details on AM-induced band splitting, and subsequently pin down its unique bulk $g$-wave symmetry through quantitative analysis of the angular and photon-energy dependence of spin splitting. Moreover, the observed spin splitting reaches the magnitude of 0.93~eV near $E_F$, the most substantial among all confirmed AM materials. This study not only validates the nature of CrSb as a prototype $g$-wave like AM material but also underscores its pivotal role in pioneering applications in spintronics.

\end{abstract}
\maketitle  

For magnetic materials, attributed to exchange interactions among atoms, spin moments align either parallel or antiparallel, giving rise to the most common ferromagnetic (FM) and antiferromagnetic (AFM) states, which were traditionally analyzed and understood primarily through magnetic group theory~\cite{brinkman1966theory,litvin1974spin,litvin1977spin,neel1971magnetism,corticelli2022spin}. Recently, researches on the spin-group-symmetry classification applicable to collinear magnets have led to the discovery of a third fundamental type of magnetism: altermagnetism (AM), characterized by distinct spin symmetry involving antiparallel sublattices that are interconnected through specific real-space rotational transformations~\cite{vsmejkal2022beyond,vsmejkal2022emerging}. Furthermore, AM exhibits 'dual-phase' magnetic behavior, displaying both FM and AFM characteristics, namely nonrelativistic spin splitting and zero net magnetization, respectively. From the perspective of spintronics device development, AM materials host unique advantages~\cite{vsmejkal2022giant,vsmejkal2022emerging,zhang2023predictable,giil2024superconductor,cheng2024orientation,fernandes2024topological}. It not only solves the problem associated with manipulating and controlling antiferromagnetic orders but also, compared to FM materials, the collinear and antiparallel magnetic configurations of AM contribute to enhanced spin dynamics and reduced sensitivity to stray magnetic fields, facilitating the development of spintronic devices~\cite{jungwirth2016antiferromagnetic,baltz2018antiferromagnetic,vsmejkal2018topological,yan2024review,chen2024emerging,ghorashi2023altermagnetic}.

\begin{figure*}[htb]
\centering
\includegraphics[width=17cm]{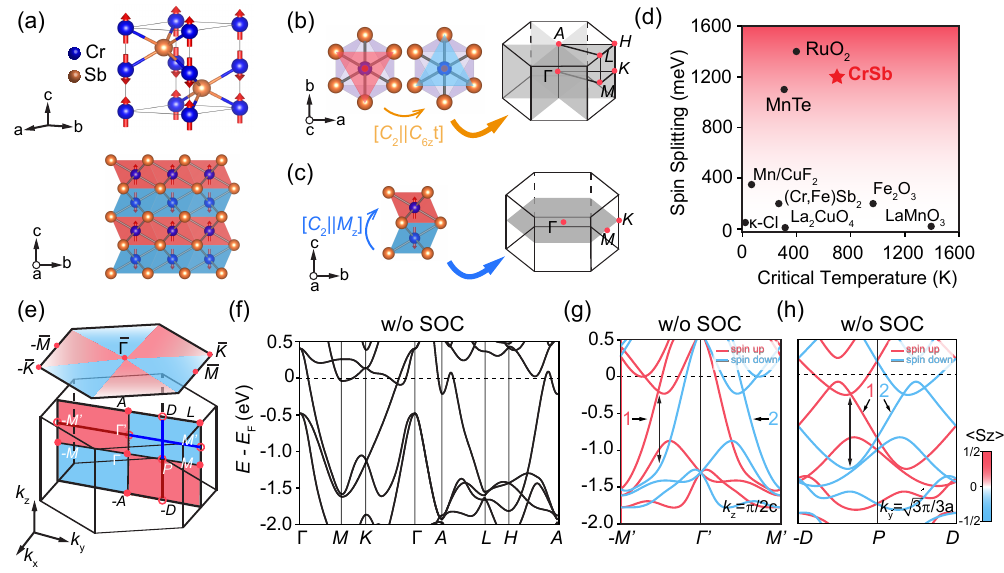}
\caption{Crystal structure and $g$-wave type large spin splitting of CrSb.
(a) Up: The unit cell of CrSb. Its spins are located at Cr-site along the $c$-axis. Down: Stack structures of spin sublattice of CrSb. 
(b-c) The opposite spin sublattices are connected via rotation and mirror operation in real space, with four nodal surfaces protected by  [$C_2 \Vert C_{6z}$$t$] and [$C_2 \Vert M_{z}$] in 3D BZ on the right panels.
(d) The magnitude of spin splitting and T$_N$ for several altermagnetic candidates. 
(e) The spin splitting distribution on the $\Gamma$$MAL$ and the projection onto (0001) plane of bands 1 and 2 corresponding to Fig.~1(g).
(f) Bulk bands without SOC along the high-symmetry paths, as determined by DFT calculation. 
(g-h) Spin splitting bands without SOC along $-M'$-$\Gamma'$-$M'$ ($k_z$ = $\pi$/2c) and $-D$-$P$-$D$ ($k_y$ = $\sqrt{3}\pi$/3$a$), respectively.
}
\label{fig1}
\end{figure*}

Theoretical predictions have suggested a diverse array of AM candidates that comply with the requisite spin group, demonstrating spin splitting characterized by $d$-, $g$-, and $i$-wave types (defined by the symmetry of spin groups~\cite{vsmejkal2022beyond,mazin2023altermagnetism}). Subsequently, researchers have reported abundant novel transport phenomena attributed to AM, such as spin polarized currents and anomalous Hall effect (AHE) in RuO$_2$~\cite{RuO2_AHE1,RuO2_AHE2,RuO2_AHE3}, spontaneous AHE in MnTe~\cite{MnTe_AHE1,MnTe_AHE2}, and both AHE and anomalous Nernst effect (ANE) in Mn$_5$Si$_3$~\cite{2024arXiv240503438S,2024arXiv240312929B}. Furthermore, recent angle-resolved photoemission spectroscopy (ARPES) studies have confirmed the band splitting and spin polarization induced by AM in the reciprocal spaces of MnTe and MnTe$_2$~\cite{lee2024broken,krempasky2024altermagnetic,zhu2024observation,osumi2024observation,hajlaoui2024temperature}. Among all AM candidates, MnTe is characterized by its collinear AFM attributes, accompanied by a $g$-wave symmetric spin texture (antisymmetric six-fold spin-momentum locking) throughout its volume, in contrast to MnTe$_2$, which is a non-collinear AFM with a distinctive plaid-like spin texture. However, the common issue of relatively small band splitting and overlapping bands in many AM materials significantly challenges the precise delineation of the three-dimensional (3D) electronic structure and identification of band splitting characteristics. In these scenarios, there remains the crucial need to identify AM candidates exhibiting substantial magnitude of spin splitting in the vicinity of Fermi level~($E_F$)~\cite{vsmejkal2022beyond,vsmejkal2022emerging}. CrSb, an isostructural sibling compound to MnTe, has emerged as a promising AM candidate, notable for its exceptionally high N$\mathrm{\acute{e}}$el temperature (T$_N$) above 700~K ~\cite{takei1963magnetic,kahal2007magnetic,tsubokawa1961magnetic,SI_Here}. Remarkably, its spin splitting in proximity to $E_F$ was suggested to be of $g$-wave symmetry and extend up to 1.2 eV, coming out on top of all predicted AM candidates (Fig.~1d), which lays the solid foundation for further application in spintronics devices~\cite{vsmejkal2022emerging,gao2023ai,zhou2024crystal}. Yet, direct high-resolution spectroscopic probing of the low-lying bands in CrSb, crucial for capturing the nuances of band splitting features near $E_F$ to definitively identify AM states, remains elusive, despite a latest soft X-ray ARPES experiment on epitaxial CrSb thin films reporting the sign of spin splitting in its electronic bands~\cite{reimers2024direct}. 

\begin{figure}[htb]
\centering
\includegraphics[width=8.5cm]{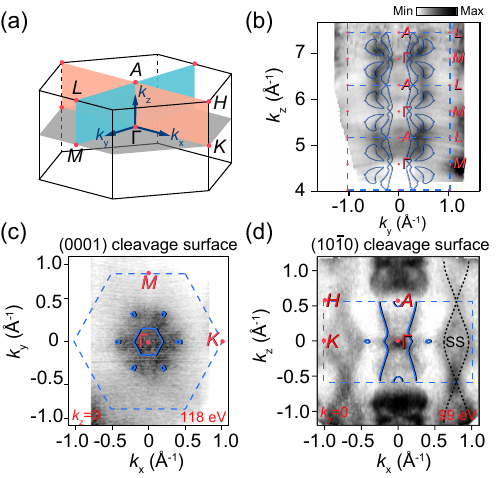}
\caption{Kramers degeneracy planes on (0001) and (10$\overline{1}$0) cleavage surfaces.
(a) A schematic diagram of three high-symmetry planes within 3D BZ. The $\Gamma$-$K$-$M$,$\Gamma$-$A$-$M$-$L$, and $\Gamma$-$K$-$A$-$H$ planes are highlighted in gray, cyan, and orange, respectively.
(b) The photon-energy dependent map of $k_z$ at $E_B$ = 0.2 eV along $\overline{\Gamma}$-$\overline{M}$ on the (0001) surface, with the constant energy contour of $\Gamma$-$M$-$A$-$L$ plane obtained from DFT calculations appended.
(c-d) Photoemission intensity maps for the $\Gamma$-$K$-$M$ plane measured at 118~eV photons and $\Gamma$-$K$-$A$-$H$ plane measured at 99~eV photons, respectively. Calculated FS are appended for reference. Surface states in the $\Gamma$-$K$-$A$-$H$ plane are marked with black dashed lines.
}
\label{fig2}
\end{figure}


\begin{figure*}[htbp]
\centering
\includegraphics[width=17cm]{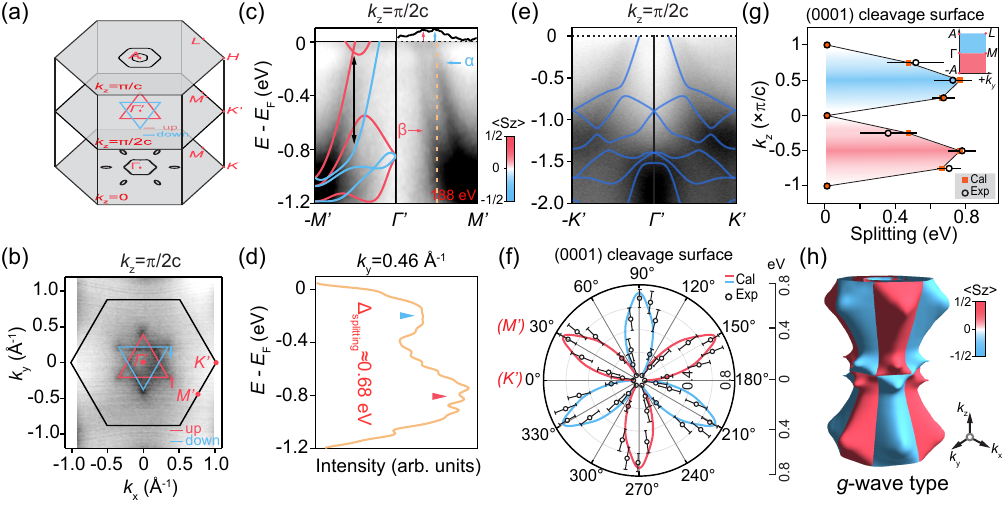}
\caption{Distribution of bands splitting in $k$ space.
(a) Calculated FS on (0001) surface at different $k_z$.
(b) FS map appended with calculated spin polarized bands at $k_z$ = $\pi$/2c (denoted as $\Gamma'$-$K'$-$M'$ plane), with red and blue representing opposite spins.
(c) ARPES intensity plot along the $-M'$-$\Gamma'$-$M'$ direction, overlaid with bulk theoretical spin splitting bands in altermagnetic state (Left half panel) and integrated momentum distribution curve (MDC) from $E_B$ = 0.1 eV to $E_F$ (Top of the right panel). 
(d) Energy distribution curve (EDC) at $k_y$ = 0.46 \AA$^{-1}$ (yellow dashed line in (c)). To enhance contrast, the Shirley background was subtracted, as detailed in the methods section in Fig.~S7.
(e) ARPES spectra overlaid with bulk theoretical bands along $-K'$-$\Gamma$-$K'$ direction.
(f) Polar distribution of the maximum spin splitting magnitude ($\Delta_{splitting}$) along the radial directions centered at $\Gamma'$ point. The radial coordinates represent the splitting size. Red and blue petals represent calculated gap size with opposite signs of spin polarization. Black hollow circles indicate the experimentally extracted $\Delta_{splitting}$ with the step of 7.5$^\circ$.  
(g) Evolution of maximum $\Delta_{splitting}$ at different $k_z$ along $\overline{\Gamma}$-$\overline{M}$ direction. The inset plot shows a spin splitting pattern along $k_z$ direction of $\overline{\Gamma}$-$\overline{M}$ path.
(h) Schematic diagram of $g$-wave type spin splitting mode in 3D $k$ space.
}
\label{fig3}
\end{figure*}

In this work, we employed synchrotron-based high-resolution ARPES in conjunction with density functional theory (DFT) calculations to present compelling evidence of the 3D electronic structure of altermagnetic features in single-crystal CrSb on both the (0001) and (10$\overline{1}$0) cleavage planes. On one hand, the spectroscopic observations are found to be in excellent agreement with DFT calculations. Significant Kramer's degeneracy of bulk bands is observed in the high-symmetry positions of momentum space, while unambiguous features of AM-induced band splitting are captured outside these high-symmetry positions. Besides, the angle-dependent AM splitting fingerprints within the Brillouin zone (BZ) confirm the $g$-wave symmetry characteristics of CrSb as a prototypical AM material. On the other hand, our photoemission spectroscopy results reveal a band splitting of up to 0.93 eV near the $E_F$, the largest observed to date among all confirmed AM materials. These findings not only highlight CrSb as an ideal candidate for investigating novel properties of AM materials but also emphasize its potential applications in spintronics.

CrSb crystallizes in a NiAs-type structure (space group P6$_3$/mmc) and exhibits an $A$-type collinear AFM ground state with the N$\mathrm{\acute{e}}$el vector aligned along $c$-axis, as illustrated in Fig.~1(a)~[The single crystal structural and magnetic characterizations see in Supplementary Notes 11 and 13~\cite{SI_Here}]. The Cr atoms, each surrounded by six Sb atoms, form two distinct sublattices with opposite spins. Figures~1(b) and 1(c) detail how the spin sublattices in CrSb are related by nonrelativistic spin-group operations [$C_2 \Vert C_{6z}$$t$] and  [$C_2 \Vert M_z$], respectively; here, $C_{6z}$$t$ represents a combined six-fold rotation and translation operator, while $M_z$ denotes the mirror operator. Significant spin splitting occurs away from these high-symmetry planes. Previous studies have predicted that MnTe, RuO$_2$, and CrSb exhibit spin splittings exceeding 1 eV~\cite{vsmejkal2022beyond,vsmejkal2022emerging}. Notably, among these materials, CrSb displays a particularly robust magnetic structure, with a $T_N$ exceeding 700~K, as summarized in Fig.~1(d). Similar to previously reported spin-momentum-locked phases in MnTe~\cite{lee2024broken}, CrSb as well presents a $g$-wave type spin splitting distribution [Fig.~1(e)]. Moreover, unlike the in-plane spin orientations in MnTe, the spins aligned along the $c$-axis in CrSb prevent the formation of domains with different N$\mathrm{\acute{e}}$el-vector orientations, providing a cleaner platform for studying the electronic structures of altermagnetism~\cite{krempasky2024altermagnetic}. Moreover, the Hall conductivities of CrSb at 2~K can be fitted using the two-band model, indicating high carrier mobility. Unlike the anomalous Hall effect (AHE) observed in MnTe~\cite{MnTe_AHE1}, our Hall measurements are absence of AHE, consistent with recent transport reports~\cite{urata2024high}. This absence of AHE is likely due to the robustness of the N$\mathrm{\acute{e}}$el vector aligned along the [001] direction under the magnetic field. Structural modulation of the Néel vector direction could be an effective approach to inducing AHE~\cite{zhou2024crystal,SI_Here}.

Our analysis begins with a comparison of low-lying band structures from high-symmetric and anti-symmetric regions in the 3D BZ. In the absence of spin-orbit coupling (SOC), the calculated bulk bands are spin-degenerate along paths $\Gamma$-$M$-$L$-$K$-$A$-$L$-$H$-$A$ [Fig.~1(f)]. However, once deviating from these high-symmetry paths, exampled with cuts $-M'$-$\Gamma'$-$M'$ ($k_z = \pi/2c$) and $-D$-$P$-$D$ ($k_y$ = $\sqrt{3}\pi$/3$a$) in Fig.~1(e), a large spin splitting of bands emerges close to $E_F$ and the corresponding $S_z$ components of split bands exhibit an anti-symmetric pattern across the high-symmetry lines, as depicted in Figs.~1(g-h) and Fig.~S1(b). We could thus summarize the spin polarization within the $\Gamma$-$M$-$L$-$A$ plane of CrSb in Fig.~1(e), in which the red (blue) denotes spin-up (spin-down), respectively. Besides, we illustrated the two-dimensional(2D) projection of spin polarization along $c$, which is reminiscent of the typical $g$-wave altermagnetism, in accordance with previous calculations~\cite{vsmejkal2022beyond,gao2023ai}. We note that our further calculations reveal an extremely subtle difference in these spin splitting with and without SOC, as depicted in Figs.~S1(c-d), suggesting the non-relativistic nature of these spin splitting.

To obtain direct photoemission evidence of the $g$-wave spin splitting distribution, it is crucial to precisely locate non-spin-degenerate momenta by means of photon-energy dependent ARPES experiments. We began by probing the (0001) cleavage surface, corresponding to the $\overline{\Gamma}$-$\overline{M}$-$\overline{K}$ panel in the reciprocal space. Our typical core-level photoemission spectra of the uneven cleavage surface suggest prominent Cr termination, with the surface states of Cr termination being suppressed~\cite{SI_Here,li2024topological}.
We conducted measurements along the $-\overline{M}$-$\overline{\Gamma}$-$\overline{M}$ path using photon energies ranging from 60 to 240 eV. The obtained $k_y-k_z$ map at $E_B$ = 0.2 eV is shown in Fig.~2(b), which corresponds to the cyan plane highlighted in Fig.~2(a). The observed periodic band dispersion closely aligns with our calculation, enabling the precise determination of the inner potential ($V_0$ = 12 eV) and thus subsequently one-to-one correspondence between the photon energy and specific $k_z$. 

Figure~2(c) presents the Fermi surface (FS) map for the $\Gamma$-$M$-$K$ plane, which was taken with 118~eV photons, complemented with the calculation that reveals an absence of spin splitting in this plane. This finding is in line with the spin degeneracy protected by the spin group symmetry [$C_2 \Vert M_z$]. To further verify all spin degeneracy along high-symmetry paths in AM, we tried to achieve the sample cleavage on the (10$\overline{1}$0) plane and then performed detailed photon-energy dependent ARPES measurements [Fig.~S3]. Through the obtained correspondence between photon energy and $k_y$, we could locate and then straightforwardly investigate the band structure in the high-symmetry $\Gamma$-$K$-$A$-$H$ plane [the orange-colored plane illustrated in Fig.~2(a)]. As shown in Fig.~2(d), FS obtained in this plane displays similar spin degeneracy across all bulk bands (dark blue lines) in the vicinity of $E_F$, which should be protected by spin group symmetry [$C_2 \Vert C_{6z}t$]. Here, we note the ribbon-like surface Fermi pockets (black dashed lines), which exhibit dispersionless features in $k_x$-$k_z$ map [Fig.~S3(c)] and are absent in our bulk band calculation.Recent researches suggest that these may be surface Fermi arcs originating from bulk band topology~\cite{lu2024observation,li2024topological}. More evidences on the spin degeneracy in other high-symmetry planes, including FS and corresponding photoemission plots have been documented in Fig.~S4.

After verifying the spin degeneracy occurring in all high-symmetry nodal planes of CrSb, we next concentrated on the quantitative identification of spin splitting away from nodal planes. We pinpointed a $\Gamma'$-$K'$-$M'$ plane that intersects the midpoint between $\Gamma$ and $A$ [illustrated in Fig.~3(a)]. The FS map at this $k_z$, depicted in Fig.~3(b), reveals a star-of-David-like structure that well matches our calculation. Furthermore, in this material AM imparts distinct spin polarization characteristics to these bands, separating them into two staggered triangular bands (marked by red and blue frames). To quantify the spin splitting, we analyzed the band dispersion along the $\Gamma'$-$M'$ direction, as shown in Fig.~3(c), where the energy offset between spin-polarized bands is most pronounced. We observed two closely parallel split bands, labeled as $\alpha$ and $\beta$, near $E_F$, gradually converging at $E_B$$\sim$1.2 eV. The integrated moment distribution curve (MDC) between $E_F$ and $E_B$ = 0.1 eV features two distinct peaks corresponding to spin-splitting bands (marked by the red and blue arrows, respectively). The calculated bulk bands for anti-symmetric spin splitting in the AM state, shown in the left panel of Fig.~3(c), align remarkably well with our ARPES data. Conversely, our spectra cannot reconcile with the calculated band structure of nonmagnetic states (Fig.~S5). Notably, compared to the vicinity of the $E_F$, the dispersion near the $\Gamma$ point around $E_B$= 1~eV exhibits weaker spectral intensity. This observation is consistent with findings from other ARPES studies and can potentially be explained by matrix element effects~\cite{SI_Here,zeng2024observation,yang2024three,li2024topological,moser2017experimentalist}. 

\begin{figure}[htbp]
\centering
\includegraphics[width=8.5cm]{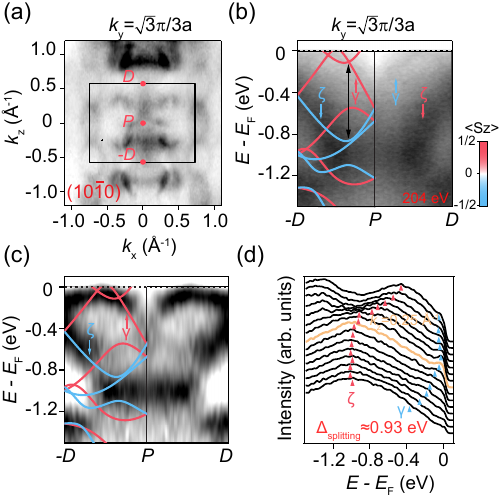}
\caption{Giant spin splitting on (10$\overline{1}$0) surface. 
(a) FS map for (10$\overline{1}$0) surface at $k_y$ = $\sqrt{3}\pi$/3$a$.
(b) Photoemission intensity plot along $-D$-$P$-$D$ direction, appended with theoretical bulk bands. Splitting bands are labeled as $\gamma$ and $\zeta$, respectively.
(c) Second derivative plot corresponding to panel (b). 
(d) EDCs within the range of $P$-$D$. The yellow curve is the EDC at $k_z$ = 0.25 \AA$^{-1}$, showing the $\Delta_{splitting}$ = 0.93 eV. 
}
\label{fig4}
\end{figure}

Attributed to high-resolution photoemission spectroscopy, we were able to accurately capture the maximum band splitting size ($\Delta_{splitting}$) of 0.68 eV at $k_y$ = 0.46 Å$^{-1}$, as determined from the peak positions of the energy distribution curve (EDC) depicted in Fig.~3(d). For comparison, we examined bands along the high-symmetry path $-K'$-$\Gamma'$-$K'$ within the same $k_z$ plane [Fig.~3(e)]. We note that these bands exhibit spin degeneracy protected by the spin group symmetry [$C_2 \Vert C_{6z}$$t$]. To better visualize the distribution of spin splitting in the reciprocal space, we present in Fig.~3(f) the distribution of $\Delta_{splitting}$ in a polar coordinate diagram centered at $\Gamma'$ points along all directions. Here, the polar angle of 0$^{\circ}$ represents the $\Gamma'$-$K'$ direction, and the polar radial axis indicates the magnitude of spin splitting. Detailed analysis for all these in-plane $\Delta_{splitting}$ along polar angles is presented in Fig.~S6. We found that the extracted $\Delta_{splitting}$ data, including their distribution and magnitude, are both in remarkable agreement with the calculation, a characteristic in-plane $g$-wave hexagonal petal. Moreover, we as well extracted $\Delta_{splitting}$ at various $k_z$ along the $\overline{\Gamma}$-$\overline{M}$ direction, as shown in Fig.~3(g). These data exhibit a periodic modulation from zero to maximum and back to zero, consistent with the theoretical $g$-wave like evolution of band splitting along $k_z$. Detailed band structure and out-of-plane $\Delta_{splitting}$ analysis are provided in Figs.~S8 and S9. On the basis of above-mentioned quantitative analysis on (0001) cleavage surface [Fig.~3(f-g)], the spin splitting pattern in 3D BZ is delineated. The bulk $g$-wave like AM existing in CrSb can be well pinned down [Fig.~3(h)]. However, it confirms that spin splitting pattern along $k_z$ direction of $\overline{\Gamma}$-$\overline{M}$ path is one-sided. Thus, studying the electronic structures of (10$\overline{1}$0) surface is necessary to provide direct evidence of the spin splitting along $k_z$ direction.   

Specifically, on the (10$\overline{1}$0) cleaved surface, we also observed spin splitting bands deviating from the high-symmetry $\Gamma$-$K$-$A$-$H$ plane. Figure~4(a) showcases a distinctive dumbbell-shaped FS located between $k_y$ = $\sqrt{3}\pi$/3$a$, which exhibits a remarkable contrast to the FS on the $\Gamma$-$K$-$H$-$A$ planes in Fig.~2(d). In Fig.~4(b), the photoemission plot along $-D$-$P$-$D$ direction is presented with appended calculated spin splitting bulk bands. 
To better capture this experimental value of $\Delta_{splitting}$, we utilized the second derivative spectra and EDCs, as shown in Fig.~4(c) and 4(d), respectively. Through these analyses, we obtained a maximum $\Delta_{splitting}$ of approximately 0.93~eV at $k_z$ = 0.25 \AA$^{-1}$, which is the record large spin band splitting observed in all reported altermagnets to date. The spin-splitting magnitude for our simple crystal is larger than the 0.6 eV reported in recent ARPES results for thin film. Such variability is anticipated given the substrate-induced compressive stress effect~\cite{reimers2024direct,yang2024three}.

In conclusion, our comprehensive study on the bulk band structure of CrSb with alternating magnetism reveals its complex spin-split electronic configuration. Utilizing high-resolution ARPES and DFT calculations, we have identified the most pronounced spin splitting of CrSb observed among altermagnets to date. 
The significant non-relativistic spin-splitting gap and spin-polarized bands intersecting the $E_F$ are predicted to generate a high spin-current ratio, making CrSb highly suitable for spintronics. The cleavage plane and momentum dependence of spin-splitting could be leveraged in future spin valves and magnetic tunnel junctions~\cite{zhou2024crystal,bai2024altermagnetism,vsmejkal2022emerging}. Our detailed analysis in three-dimensional momentum space demonstrates a unique $g$-wave type spin pattern in CrSb, highlighting its potential as a platform for investigating novel phenomena in the interplay of electronic and magnetic correlations.

This work was supported by National Key R\&D Program of China (Grants No. 2023YFA1406304 and 2023YFA1406100), National Science Foundation of China (Grant Nos. U2032208, 12004405) and the New Cornerstone Science Foundation. Y. F. Guo acknowledges the open research funds of State Key Laboratory of Materials for Integrated Circuits (Grant No. SKL2022) and Beijing National Laboratory for Condensed Matter Physics (2023BNLCMPKF002). Y.L. Wang was supported by the Innovation Program for Quantum Science and Technology (No. 2021ZD0302800) and the National Natural Science Foundation of China (No. 12174365). Z. C. Jiang acknowledges the China National Postdoctoral Program for Innovative Talents (BX20240348). J. P. Sun and J. G. Cheng acknowledge the National Natural Science Foundation of China (12025408, 12174424). Part of this research used Beamline 03U of the Shanghai Synchrotron Radiation Facility, which is supported by ME$^2$ project under contract No. 11227902 from National Natural Science Foundation of China.

\bibliographystyle{naturemag}
\bibliography{CrSb}
\end{document}